\documentclass[%
 reprint,
superscriptaddress,
 amsmath,amssymb,
 aps,
 pra,
]{revtex4-2}

\usepackage{graphicx}
\usepackage{dcolumn}
\usepackage{bm}
\usepackage[colorlinks,linkcolor=blue,anchorcolor=blue,citecolor=blue,urlcolor=blue]{hyperref}
\usepackage{xcolor}

\usepackage{verbatim}

\begin{document}

\title{Waveguide QED with dissipative light-matter couplings}

\author{Xing-Liang Dong}
\affiliation{Ministry of Education Key Laboratory for Nonequilibrium Synthesis and Modulation of Condensed Matter, Shaanxi Province Key Laboratory of Quantum Information and Quantum Optoelectronic Devices, School of Physics, Xi’an Jiaotong University, Xi’an 710049, China }
\affiliation{Theoretical Quantum Physics Laboratory, Cluster for Pioneering Research, RIKEN, Wakoshi, Saitama 351-0198, Japan}

\author{Peng-Bo Li}
\email{lipengbo@mail.xjtu.edu.cn}
\affiliation{Ministry of Education Key Laboratory for Nonequilibrium Synthesis and Modulation of Condensed Matter, Shaanxi Province Key Laboratory of Quantum Information and Quantum Optoelectronic Devices, School of Physics, Xi’an Jiaotong University, Xi’an 710049, China }

\author{Zongping Gong}
\affiliation{Department of Applied Physics, The University of Tokyo, 7-3-1 Hongo, Bunkyo-ku, Tokyo 113-8656, Japan}
\affiliation{Theoretical Quantum Physics Laboratory, Cluster for Pioneering Research, RIKEN, Wakoshi, Saitama 351-0198, Japan}

\author{Franco Nori}
\affiliation{Theoretical Quantum Physics Laboratory, Cluster for Pioneering Research, RIKEN, Wakoshi, Saitama 351-0198, Japan}
\affiliation{Center for Quantum Computing, RIKEN, Wakoshi, Saitama 351-0198, Japan}
\affiliation{Physics Department, The University of Michigan, Ann Arbor, Michigan 48109-1040, USA }

\begin{abstract}

Dissipative light-matter coupling plays a vital role in non-Hermitian physics, but it remains largely unexplored in waveguide QED systems.
In this work, we find that by employing pseudo-Hermitian symmetry rather than anti-PT symmetry, the concept of dissipative coupling could be generalized and applied to the field of waveguide QED. This leads to a series of intriguing results, such as spontaneous breaking of pseudo-Hermitian symmetry across the exceptional points (EPs), level attraction between the bound states, and critical transition across the EPs for the population of quantum emitters in the bound state. Thanks to the tunability of photonic bands in crystal waveguides, we also demonstrate that dissipative light-matter coupling leads to the emergence of nonstandard third-order exceptional points with chiral spatial profiles in a topological waveguide QED system. This work provides a promising paradigm for studying non-Hermitian quantum phenomena in waveguide QED systems.

\end{abstract}

\date{\today}

\maketitle


\emph{Introduction---.}
Waveguide quantum electrodynamics (QED) is a well-known field in quantum optics,
where light confined in one-dimension interacts with quantum emitters \cite{RevModPhys.95.015002,RevModPhys.90.031002}.
In this field, photonic structures with bandgaps are promising
for tailoring nontrivial light-matter interactions and advancing quantum information processing and simulation
\cite{PhysRevLett.64.2418,PhysRevLett.115.063601,Quantum2015Douglas,
Barik666,PhysRevResearch.2.023003,PhysRevLett.126.203601,PhysRevA.106.043703,
Sipahigil847,Belloeaaw0297,PhysRevLett.124.213601,PhysRevX.11.011015,
Xue2023A,PhysRevLett.131.073602}.
Another idea is based on directly manipulating light-matter couplings,
which can be nonlocal \cite{PhysRevLett.120.140404,Kannan2020Waveguide,PhysRevLett.126.043602},
ultrastrong \cite{PhysRevResearch.4.023194,PhysRevA.106.063717,PhysRevA.109.033715},
and nonlinear \cite{PRXQuantum.4.030326}.
Recently, introducing non-Hermiticity into structured waveguide QED
has revealed a large amount of novel non-Hermitian quantum phenomena \cite{PhysRevA.93.062129},
including, but not limited to, hidden bound states (BSs) with skin-effect origin
\cite{PhysRevLett.129.223601,PhysRevA.106.053517,ro2024hermitian},
fractional quantum Zeno effect \cite{PhysRevX.13.031009}
and nonreciprocal light-mediated interactions \cite{Federico2022Exotic,PhysRevResearch.5.L042040}.
Despite these tremendous progresses, previous studies mainly focus on quantum emitters
coherently coupled to photonic waveguide structures with engineered complex energy band structures. That is, the photonic waveguide is by itself a non-Hermitian bath, leaving the problem of dissipative emitter-photon waveguide coupling unexplored.

On the other hand, dissipative couplings play a crucial role in non-Hermitian physics
\cite{Yuto2020Non,RevModPhys.93.015005,Qiang2023Non,PhysRevResearch.5.L042005,
dreon2022self,leefmans2022topological}, typically in a two-mode system coupled to an auxiliary dissipative reservoir which induces a dissipative intermode coupling \cite{PhysRevX.5.021025}. In this context, it features experimentally observable level attraction
\cite{OKOLOWICZ2003271,PhysRevLett.121.137203,PhysRevLett.128.047701,Wurdack2023Negative,PhysRevA.109.023707}
and anti-parity-time (anti-PT) symmetric phase transitions
\cite{Peng2016Anti,PhysRevA.96.053845,Choi2018Observation,Ying2019Anti,Zhang2020Breaking,
PhysRevLett.126.180401,Yang2022Radiative,PhysRevLett.131.103602}.
The latter is associated with exceptional points (EPs), where both eigenvalues and the associated eigenstates coalesce
\cite{Ozdemir2019Parity,Huang2022Exceptional}.
These have potential applications, such as PT-symmetric lasers
\cite{PhysRevLett.113.053604,Liang2014Single,Hossein2014Parity},
and EP-enhanced sensing \cite{Hodaei2017Enhanced,Chen2017Exceptional}.
However, there lacks a comprehensive analysis of dissipative coupling of quantum emitters to a photonic continuum with extensive degrees of freedom. This may possess a number of
surprising and potentially useful features that have never been discovered before in few-mode systems.

In this work, we address these critical yet unexplored problems by investigating the non-Hermitian aspects of dissipative light-matter couplings in a waveguide QED system.
We find that this dissipative waveguide QED system has pseudo-Hermitian symmetry
\cite{Mostafazadeh2002Pseudo,Mostafazadeh2002Pseudo2} rather than anti-PT or PT symmetry prevalent in the few-mode case. This symmetry describes the Hamiltonian as effectively Hermitian
via an invertible linear operator $\eta$, i.e., $\eta H_{pH}\eta^{-1}=H_{pH}^\dag$. It is capable of producing real eigenvalues and undergoing spontaneous breaking across EPs. Different from previous studies with bandgap structures \cite{PhysRevLett.101.100501,PhysRevA.93.033833}, here we find that the bound states of this pseudo-Hermitian system display level attraction, and more interestingly, a critical transition across  the EPs for the QE population.
In the pseudo-Hermitian symmetry broken phase, surprisingly, we find the QE population in the formed bound state is always fixed as $1/2$. In the unbroken phase, the
single-QE dynamics displays prominent population retrieval on top of fractional decay.
In addition, we find intriguing phenomena in a topological waveguide QED system
\cite{Belloeaaw0297,PhysRevX.11.011015},
including the emergence of nonstandard third-order EPs
\cite{Demange_2012,PhysRevLett.127.186601,PhysRevA.104.063508,
PhysRevResearch.4.023130,PhysRevB.108.115427} and chiral bound states.
This work initiates the study of dissipative coupling
to a photonic continuum with tunable band structures,
thus opening a new avenue for exploring non-Hermitian physics in waveguide QED.

\begin{figure}[t]
\includegraphics[scale=0.14]{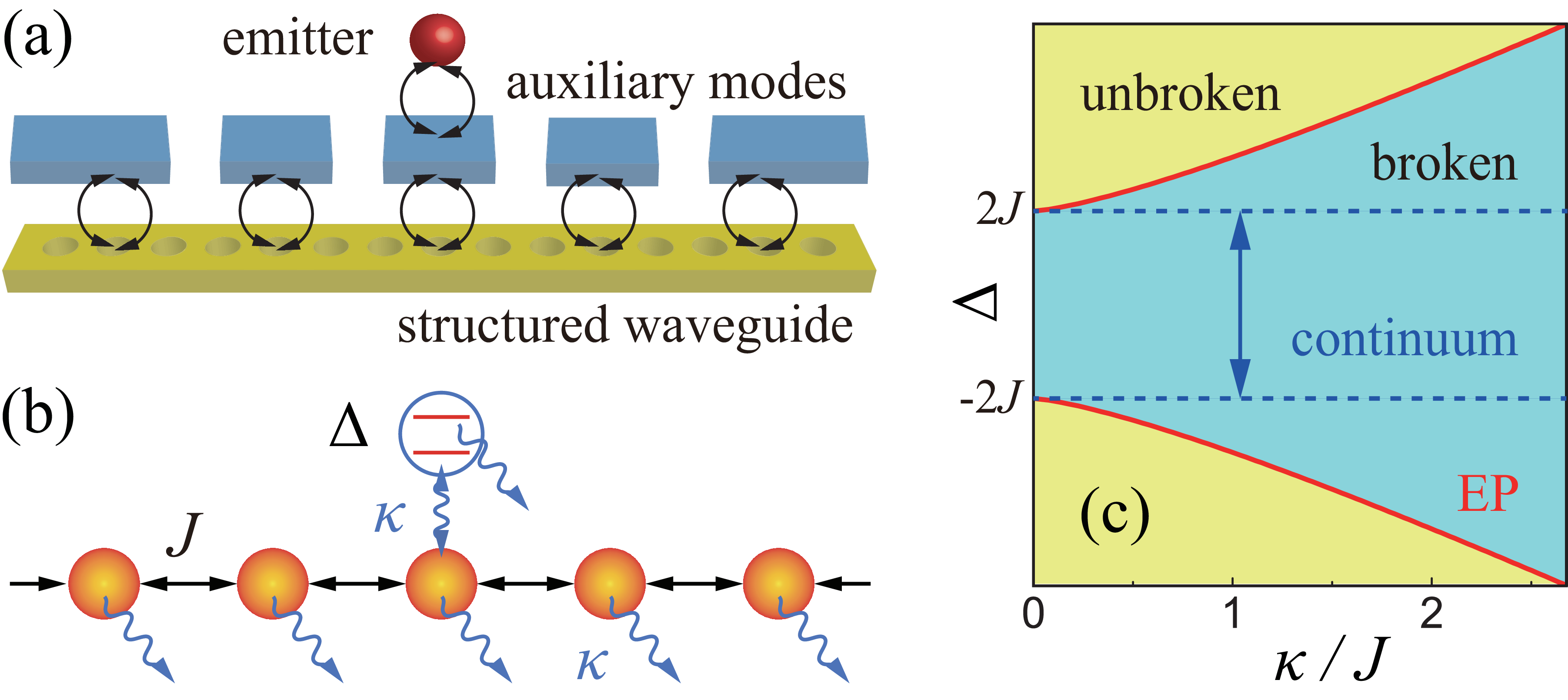}
\caption{\label{fig1}
(a) Schematic illustration of a two-level emitter dissipatively coupled to
a 1D structured bath through auxiliary modes.
(b) The reduced model of one emitter in a coupled-resonator waveguide after adiabatically eliminating the intermediate modes. Typically, decay into neighboring auxiliary modes is negligible, validating the assumption of independent decay channels \cite{PhysRevX.11.011015,Xue2023A}.
(c) Single-excitation phase diagram of the pseudo-Hermitian waveguide QED system (b).}
\end{figure}

\emph{Model and Hamiltonian.---}
We consider the setup where two-level QEs are dissipatively coupled to a tight-binding resonator chain; this dissipative coupling
can be realized with the help of auxiliary modes, as depicted in Figure 1.
In the Markovian limit, the dissipative dynamics can be described by a Lindblad master equation (let $\hbar=1$)
\begin{eqnarray}\label{ME1}
\dot{\hat\rho}&=&-i[\hat{H}_B+\hat{H}_S,\hat\rho]\notag\\
&+&2\kappa\sum_m\mathcal{D}[\hat{a}_{n_m}+\hat\sigma_m]\hat\rho
+2\kappa\sum_{n\neq n_m}\mathcal{D}[\hat{a}_{n}]\hat\rho,
\end{eqnarray}
where $\hat{H}_B=-J\sum_{n}(\hat{a}_{n+1}^\dag\hat{a}_n+\text{H.c.})$
is the Hamiltonian of the 1D photonic lattice with hopping strength $J$ and annihilation operator $\hat{a}_n$ on site $n$,
$\hat{H}_S=\Delta\sum_m\hat\sigma_m^\dag\hat\sigma_m$
is the free Hamiltonian of the quantum emitters with detuning $\Delta$ and spin operator $\hat\sigma_m^\dag=|e\rangle_m\langle g|$,
and $\mathcal{D}[\hat{L}]\hat\rho\equiv\hat{L}\hat\rho\hat{L}^\dag-\{\hat{L}^\dag\hat{L}/2,\hat\rho\}$
is the Lindblad superoperator.
The second term on the right gives rise to a dissipative light-matter interaction $\hat{H}_\text{int}=-i\kappa\sum_m(\hat{a}_{n_m}^\dag\hat\sigma_m+\text{H.c.})$,
along with local dampings on coupled sites $\hat{a}_{n_m}$ and emitters $\hat\sigma_m$ with identical strength $\kappa$.
The last term is added to achieve a uniform global on-site photon loss.

Background loss shifts the diagonal imaginary energy without affecting phase transitions and EPs.
The impact on the dynamics appears as a time decay of $e^{-\kappa t}$,
which can be mitigated under $\text{Tr}(\hat\rho)=1$.
This method is widely used in PT-symmetric experiments with a passive system replacing an active one
\cite{PhysRevLett.103.093902,PhysRevLett.108.024101,doi:10.1126/science.1258004}.
For simplicity, we focus on the effective non-Hermitian Hamiltonian, $\hat{H}_\text{eff}=\hat{H}_B+\hat{H}_S+\hat{H}_\text{int}$.
This is justified by using designed post-selection schemes that select trajectories without quantum jumps \cite{bender2007making,Naghiloo2019Quantum,PhysRevA.101.062112,harrington2022engineered}.
Alternatively, inspired by anti-PT symmetric systems \cite{PhysRevLett.125.147202}, we suggest extending this approach to continuous modes to retain the transmission spectrum as a useful observable
\cite{SupplementalMaterial}.

Here, we do not claim that the dissipative coupling itself is novel. Instead, our main message is that there remain unexpectedly rich phenomena to be explored at the interface between non-Hermitian physics and waveguide QED. This framework allows us to explore unique phenomena, such as level attraction, critical transitions, and symmetry breaking across the EPs, which extend beyond the scope of traditional quantum optical studies. Note that we can definitely go to the classical regime. The problem is inspired by waveguide QED, yet it is not restricted to waveguide QED.

By performing the Fourier transformation $\hat{a}_k=N^{-1/2}\sum_ne^{-ikn}\hat{a}_n$,
we can transform the effective non-Hermitian Hamiltonian into
\begin{eqnarray}\label{ME2}
\hat{H}_\text{eff}&=&\hat{H}_S+\sum_{k}\omega_k\hat{a}_k^\dag\hat{a}_k\notag\\
&-&i\kappa N^{-1/2}\sum_m\sum_k(e^{-ikn_m}\hat{a}_k^\dag\hat\sigma_m+\text{H.c.}),
\end{eqnarray}
with dispersion relation $\omega_k=-2J\cos k$ and lattice period $d_0=1$.
The system's nonunitary dynamics is governed by
\cite{PhysRevLett.109.230405}
\begin{eqnarray}\label{ME3}
\dot{\hat\rho}=-i[\hat{H}_B+\hat{H}_S,\hat\rho]-i\{\hat{H}_\text{int},\hat\rho\}
+2i\text{Tr}(\hat\rho\hat{H}_\text{int})\hat\rho,
\end{eqnarray}
where the third term is added to preserve probability conservation. For a pure state, this can be simplified as $e^{-i\hat{H}_\text{eff}t}|\psi_0\rangle/\sqrt{\langle \psi_t | \psi_t \rangle}$.
Focusing on the QEs subsystem, one can derive the excited-state population dynamics
of an arbitrary emitter using $\langle\hat\sigma_m^\dag(t)\hat\sigma_m(t)\rangle=\langle e_m,0|\hat\rho(t)|e_m,0\rangle$.
Unlike conventional open systems that consider steady-state experiments with a drive,
we emphasize more on transient dynamics.

\emph{Pseudo-Hermitian waveguide QED.---}
In this work, we restrict our study to the single-excitation sector
\cite{PhysRevLett.118.200401,PhysRevA.96.043811}.
We start by considering a single emitter.
For $J=0$, anti-PT symmetry can be restored
by shifting the real energy by $\Delta/2$,
a result that has been extensively studied (e.g., two atoms coupled to a photon reservoir)
\cite{Peng2016Anti,PhysRevLett.128.173602}.
However, when $J\neq0$, a photonic continuum forms and anti-PT symmetry is not applicable.

\textcolor{black}{
To address this issue, we employ pseudo-Hermitian symmetry $\hat\eta\hat{H}_\text{eff}\hat\eta^{-1}=\hat{H}_\text{eff}^\dag$,
where $\hat\eta=\hat\sigma^\dag\hat\sigma-\sum_{k}\hat{a}_k^\dag\hat{a}_k$
\cite{SupplementalMaterial}.
This implies that having identical on-site loss is crucial,
as it enables the exploration of pseudo-Hermitian physics in lattice models
\cite{Mostafazadeh2002Pseudo,Mostafazadeh2002Pseudo2}.
Note that expanding into the waveguide continuum introduces
fundamentally different subjects,
such as band structures and QE-photon bound states.
This foundation is further enriched with band topology,
vacancy-like dressed states
\cite{PhysRevLett.126.063601}, and long-range tunable interactions.}

The single-QE phase diagram of such a pseudo-Hermitian waveguide QED system is plotted in Fig.~1(c),
with entirely real eigenenergies (one complex-conjugate pair) in the unbroken (broken) phase.
When the detuning is inside the band,
the pseudo-Hermitian symmetry is broken unless the coupling is  zero.
In contrast, a phase transition at the EP is allowed outside the continuum.

\begin{figure}[t]
\includegraphics[scale=0.18]{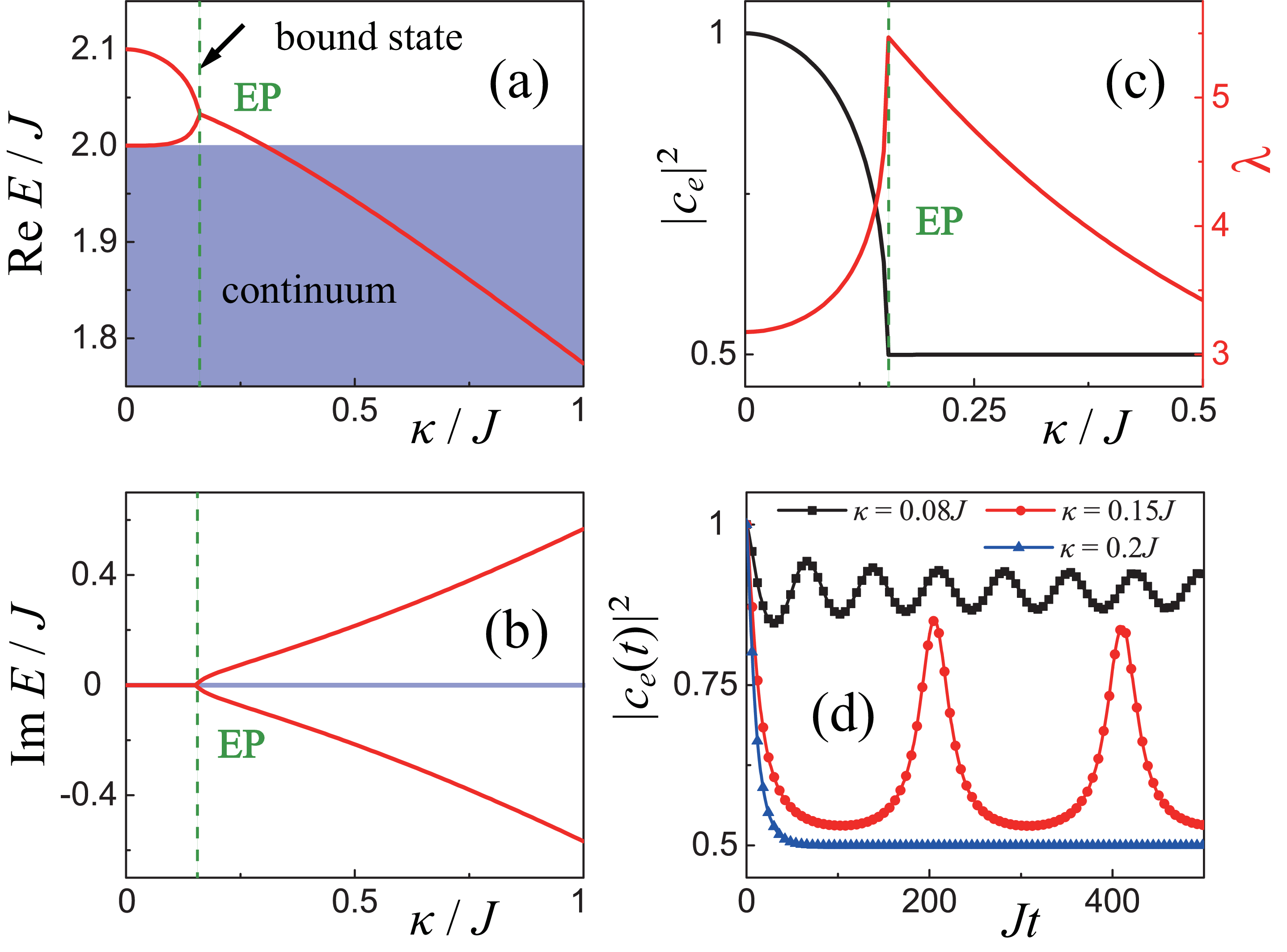}
\caption{\label{fig2} (a,b) Complex energy spectrum of one emitter in the waveguide,
with $\Delta=2.1J$.
(c) Population $|c_e|^2$ and localization length $\lambda$
as a function of coupling strength for the upper BS shown in (a).
(d) Excited-state population dynamics under different coupling strengths,
with $\kappa_\text{EP}\sim0.16J$.}
\end{figure}

\emph{Level attraction and phase transition.---}
A natural question arises: how do the non-Hermitian phenomena manifest
in pseudo-Hermitian waveguide QED?
Let us first consider that the QE's frequency lies inside the upper bandgap $\Delta>2J$.
In this regime, the BSs localized around emitters appear naturally. In Figs.~2(a,b),
level attraction and a phase transition are observed in the BSs,
with spectral singularity determined by $E_{ep}=(\Delta+\sqrt{\Delta^2+32J^2})/4$ [based on Eqs.~(\ref{ME4},\ref{ME5})].
A corresponding transition of the population $|c_e|^2$
and localization length $\lambda$ is shown in Fig.~2(c).
\textcolor{black}{The parameter $\lambda$ describes the wave function diminishes with the distance as $e^{-n/\lambda}$.
Unlike conventional scenarios,
increased coupling strength may lead to a larger localization length.}

This indicates a  phase transition in the emitter dynamics across the EP. As shown in Fig.~2(d),
in the unbroken phase, the single-QE dynamics displays prominent population retrieval
on top of fractional decay; in the broken phase, the QE population in the formed bound state
always keeps the value of 1/2.
Similar phenomena have been observed in PT-symmetric systems
\cite{PhysRevLett.119.190401,PhysRevLett.123.230401}, but here we uncover these phenomena in a fundamentally different setting.
We also numerically compute the two-QE dynamics (see Ref.~\cite{SupplementalMaterial}),
which is qualitatively different in the two phases.

\emph{Bound states in the broken phase.---}
We now consider the band regime, where $\Delta\in[-2J,2J]$,  corresponding to the pseudo-Hermitian symmetry broken phase.
In the weak-coupling limit ($\kappa\ll J$),
the Fermi's golden rule predicts a negative single QE's decay rate
$\Gamma(k)=(-i\kappa)^2/|v_g|<0$,
with $v_g=\partial\omega_k/\partial k$ the photon group velocity.
This anomaly implies the appearance of a pair of bound states with complex-conjugate eigenenergies $E_b=E_\pm$.
\textcolor{black}{
An intuitive explanation is that the interaction of the QE with the resonant mode $k(\Delta)$
breaks pseudo-Hermitian symmetry, thus forming a QE-photon bound state.}
In Fig.~3(a), we plot the real-space photon profile of the bound states when $\Delta=0$
and find a relatively large localization length $\lambda\sim44.46$ for $\kappa=0.3J$.
This is much larger than that of the bound state formed in the bandgap regime by replacing $\Delta=2.1J$.
We also show that the localization length is divergent close to $\kappa=0$
and decays exponentially with the coupling strength in Fig.~3(b).

\begin{figure}[t]
\includegraphics[scale=0.18]{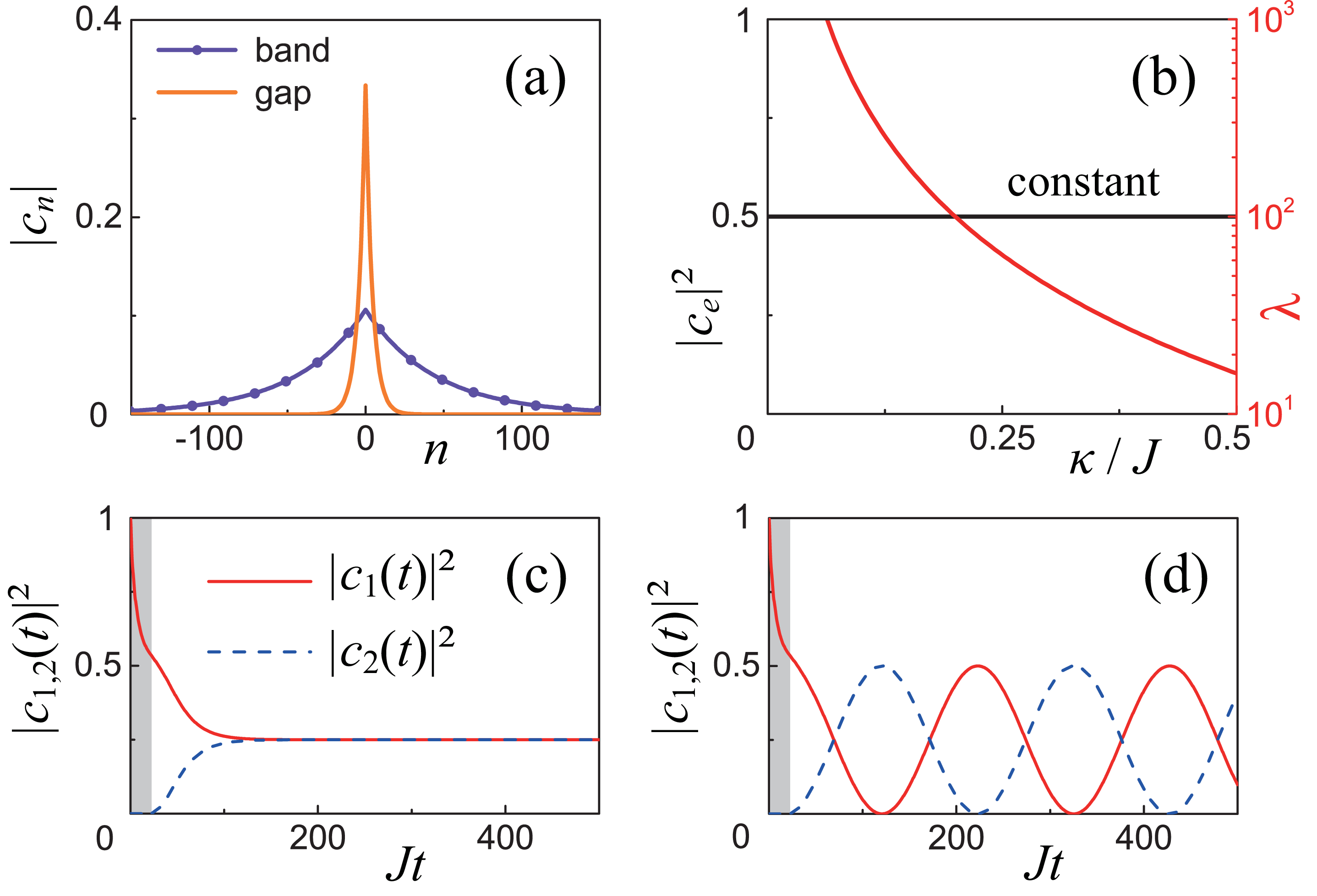}
\caption{\label{fig3}
(a) Photon profiles of the BSs,
with $\Delta=0$ ($\Delta=2.1J$) and $\kappa=0.3J$ for the band (gap) case.
(b) Population $|c_e|^2$ and localization length $\lambda$
versus the coupling strength for BSs formed in the band regime.
(c,d) Two-QE population dynamics at zero detuning,
with (c) even emitter separation $n_{12}=40$ and (d) odd $n_{12}=41$.}
\end{figure}

Specifically, we solve $\hat{H}_\text{eff}|\psi\rangle=E_b|\psi\rangle$
with single-excitation wave function
$|\psi\rangle=[c_e\hat{\sigma}^\dag+N^{-1/2}\sum_kc_k\hat{a}_k^\dag]|g\rangle|0\rangle$,
which yields $\Delta c_e-(i\kappa/N)\sum_kc_k=E_bc_e$, and $\omega_kc_k-i\kappa c_e=E_bc_k$, $\forall k$.
We assume the single QE couples to the $n_0=0$ site.
Then, the eigenvalues are given by the pole equation
\begin{eqnarray}\label{ME4}
E_b-\Delta-\Sigma_e(E_b)=0,\quad \Sigma_e(E)=-\frac{\kappa^2}{N}\sum_{k}\frac{1}{E-\omega_k},
\end{eqnarray}
where $\Sigma_e(E)$ is the self-energy of the QE.
In the thermodynamic limit $N\to\infty$, $\Sigma_e(E)=-\kappa^2/(E\sqrt{1-4J^2/E^2})$.
In particular, analytic intuition can be obtained at zero detuning, where
$E_\pm=\pm\sqrt{2J^2-\sqrt{4J^4+\kappa^4}}$ are purely imaginary.
The photon profile can be calculated by means of the inverse Fourier transform
$c_n=N^{-1/2}\sum_ke^{ikn}c_k$.
The  population of the QE can be obtained from the normalization condition
\begin{eqnarray}\label{ME5}
|c_e|^2=(1+\kappa^2N^{-1}\sum_{k}|E_b-\omega_k|^{-2})^{-1},
\end{eqnarray}
and $c_e\equiv1/\sqrt{2}$ [see Fig.~3(b)],
which corresponds to a QE-photon dressed state with equal population weights.
This can be derived by combining Eq.~(\ref{ME4})
and its complex conjugate and Eq.~(\ref{ME5}), leading to
\begin{eqnarray}\label{ME6}
(E_b-E_b^*)(2-|c_e|^{-2})=0.
\end{eqnarray}
In the broken phase, $E_b\neq E_b^*$, therefore we have $2-|c_e|^{-2}=0$,
\textcolor{black}{
aligning with Bell-like states in anti-PT systems
\cite{PhysRevLett.124.053602,PhysRevB.106.L180406}.}
This conclusion can be generalized to arbitrary two Hermitian subsystems with purely dissipative couplings
\cite{SupplementalMaterial}, which naturally covers the multi-QE case, i.e., $\sum_{m}|c_m|^2\equiv1/2$.

\emph{Two-QE dynamics at $\Delta=0$.---}
Now we focus on the case where two QEs are coupled to band of modes on resonance. In the broken phase, the bound states with eigenenergies of maximal imaginary parts dominate the long-time dynamics.
For most $\Delta\in[-2J,2J]$, there are two pairs of complex-conjugate solutions
$E_b=E_{\pm,p}$ ($p=s,a$) with unequal imaginary values.
\textcolor{black}{Only the bound states with eigenenergy satisfying $\text{Im}E_{+,p}>0$
may contribute to the dynamics after a certain period of evolution,
ultimately leading to a dominant state.
In particular, when $\text{Im}E_{+,s}\approx\text{Im}E_{+,a}$,
both bound states are reserved in the long-time limit,
and long-lived oscillatory behavior between the bound states can be observed \cite{SupplementalMaterial}.}

Choosing $\Delta=0$ and the emitter separation $n_{12}=n_2-n_1$ being an even number,
one of the complex-conjugate pairs reduces to a zero-energy solution, leaving a single bound state as the final state of the system in the long-time limit [see Fig.~3(c)].
\textcolor{black}{
As for an odd $n_{12}$, $E_{+,s}=-E_{+,a}^*$ and the QE's population dynamics is well captured by
$|c_1(t)|^2=1/2\cos^2[\text{Re}(E_{+,s})t]$ and $|c_2(t)|^2=1/2\sin^2[\text{Re}(E_{+,s})t]$,
exhibiting a perfect coherent transfer of the emitter excitation, as plotted in Fig.~3(d).
The sum of the populations satisfies $|c_1(t)|^2 + |c_2(t)|^2 = 1/2$,
consistent with the behavior of BSs in the broken phase.
The separation-dependent dynamics is closely related to interference effects,
resembling the photon-mediated collective interactions
in an array of emitters in a vacuum with emitter spacing $d=n\lambda_0/4$, where $\lambda_0$ is the photon wavelength.
In the Hermitian case, the dipole-dipole interactions mediated
by the BSs also demonstrate incomplete energy exchanges
\cite{PhysRevA.87.033831,PhysRevX.9.011021};
however, the population on QEs decreases monotonically with increasing coupling strength, different from the results in this work.}

\begin{figure}[t]
\includegraphics[scale=0.19]{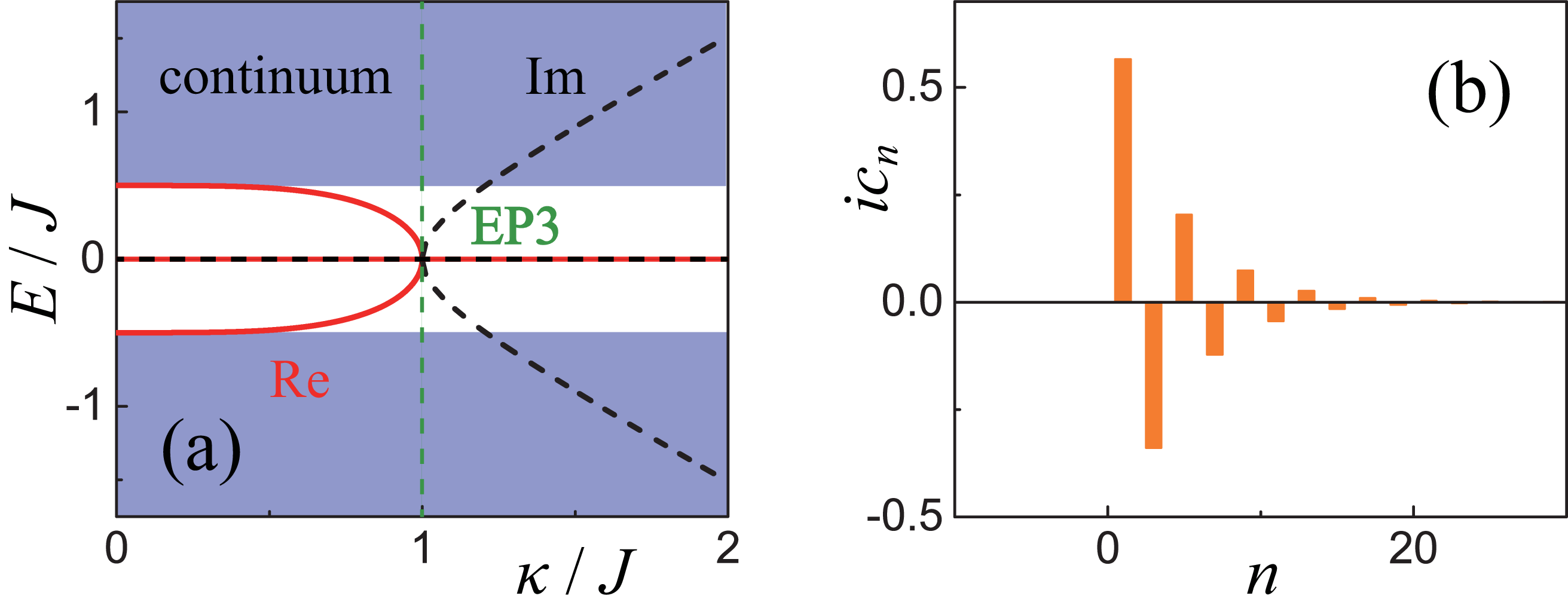}
\caption{\label{fig4}
Emitters in a topological waveguide.
(a) Complex energy spectrum of one emitter dissipatively
coupled to $A$ sublattice at the $j_0=0$ unit cell.
Here, $\Delta=0$ and $\delta=0.25$.
(b) Photon profile of the eigenstate at the third-order EP.}
\end{figure}

\emph{Extension to a topological waveguide.---}
The photonic bands exhibit high tunability,
offering the potential for more intriguing phenomena.
As an example, we consider the SSH chain.
In terms of $\hat{V}^\dag_k=(\hat{a}^\dag_k,\hat{b}^\dag_k)$,
where $\hat a_k$/$\hat b_k$ annihilates a photon with momentum $k$ in sublattice $A$/$B$,
the topological lattice Hamiltonian is written as
$\hat{H}_\text{SSH}=\sum_k\hat{V}^\dag_kh_k\hat{V}_k$, with the Bloch Hamiltonian
\begin{eqnarray}\label{ME7}
h_k=-J[(1+\delta)+(1-\delta)\cos k]\sigma_x-J(1-\delta)\sin k\sigma_y,
\end{eqnarray}
where $\delta$ is the dimerization parameter.
The $\Delta=0$ regime is of particular interest,
in which the bulk topology can be reflected in the edge modes.

\textcolor{black}{Introducing topology into the waveguide QED results in
the emergence of a special class of BSs, known as vacancy-like dressed states
\cite{PhysRevLett.126.063601},
whose eigenenergy does not depend on the coupling strength.}
In Fig.~4(a), we plot the energy spectrum for a single QE
coupled to sublattice $A$ at the $j_0$th unit cell, with $\delta>0$.
It shows that the zero detuning seeds three BSs: one vacancy-like dressed state
and two from the inner band edges with eigenenergies
\begin{eqnarray}\label{ME8}
E_\pm^\text{SSH}=\pm\sqrt{2J^2(1+\delta^2)-\sqrt{4J^4(1-\delta^2)^2+\kappa^4}}.
\end{eqnarray}
When $\kappa=2J\sqrt{\delta}$, these three BSs coalesce at a third-order EP, with the eigenstate being
a maximal superposition of the QE excited state and the topological edge state
$|\psi_\text{vds}^{j_0,A}\rangle=\big(|e\rangle|\text{vac}\rangle-i|g\rangle|\text{ES}\rangle\big)/\sqrt{2}$,
where $|\text{vac}\rangle$ denotes the photon vacuum and
$|\text{ES}\rangle=\sum_{j\geq 0}2\sqrt{\delta}(\delta-1)^j/(1+\delta)^{j+1}\hat{b}_{j+j_0}^\dag|\text{vac}\rangle$,
with unit cell index ($j+j_0$).
Note that the nonstandard third-order EP, instead of a second-order EP plus a normal mode,
asymptotically has a square root dispersion $\sim\kappa^{1/2}$
[cf.~Eq.~(\ref{ME8})] due to the sublattice symmetry
\cite{Demange_2012,PhysRevLett.127.186601,PhysRevA.104.063508,PhysRevResearch.4.023130,PhysRevB.108.115427}.
\textcolor{black}{
In a manner similar to the passive role of the flat band in a non-Hermitian three-band system
\cite{PhysRevLett.127.186601},
the vacancy-like dressed state in dissipative topological waveguide QED
is utilized to lift the order of the EP.}
The photonic part of the dressed state with a chiral profile is plotted in Fig.~4(b),
showing that the wave function appears to the right side of the QE
and exclusively on the sublattice $B$.

\textcolor{black}{
Third-order EPs may offer enhanced sensitivity in sensing applications
\cite{Hodaei2017Enhanced},
but their study in waveguide QED is largely unexplored.
The emergence of the third-order EPs in the SSH chain
showcases the interplay of bound state, topology, and pseudo-Hermiticity,
providing fresh insights into the formation of these singularities.
When applied to multiple QEs, the BSs with chiral profiles can not only generate chiral interactions
\cite{Belloeaaw0297},
but also give rise to additional novel phenomena without Hermitian analogs.
These include switchable light-mediated interactions,
and long-time dynamics dominated by the degenerate EPs \cite{SupplementalMaterial}.}

\emph{Discussion and conclusion.---}
1D resonator waveguides, including those with topological structures,
have already been demonstrated in various experimental platforms,
such as superconducting circuits \cite{PhysRevX.11.011015,Xue2023A},
nanocavities \cite{Photonic2022Saxena} and optomechanical lattices \cite{Youssefi2022Topological}.
\textcolor{black}{For example, we can build a lattice model using superconducting metamaterials made of capacitively coupled arrays of $LC$ resonators. Each $LC$ resonator is connected to a transmission line to introduce on-site dissipation. Additionally, superconducting transmon qubits can be connected to transmission lines to create dissipative light-matter interactions \cite{SupplementalMaterial}. In circuit QED experiments \cite{murch2013observing,Naghiloo2019Quantum,PhysRevLett.131.260201}, post-selection is done by reconstructing the full density matrix of the quantum state using quantum state tomography. During data processing, only the matrix elements within the $n$-excitation sector (e.g., the single-excitation sector) are kept, while others are set to zero. }

In summary, we have studied QEs dissipatively coupled to structured waveguides.
This is a simple, yet unexplored non-Hermitian generalization of the conventional
light-matter interacting systems.
By identifying pseudo-Hermitian symmetry instead of anti-PT
symmetry as the essential underlying symmetry,
we have carried out a comprehensive analysis of dissipative coupling to a photonic continuum.
Introducing pseudo-Hermiticity into waveguide QED revealed several remarkable results,
such as level attraction between the BSs and critical transition across the EPs for the QE population.
The approach is universal and adaptable to various band structures,
even extending to the multi-excitation sector
\cite{PhysRevX.6.021027,SupplementalMaterial}.
When applied to the SSH lattice, we found nonstandard third-order EPs.
Finally, we point out that one may develop
practical quantum applications based on this new paradigm,
harnessing the unique properties of non-Hermitian systems for technological advancements.

We acknowledge Yexiong Zeng, Xuefeng Pan, Peirong Han, Wei Qin for helpful discussions.
The simulation is carried out using the QuTiP \cite{JOHANSSON20121760,JOHANSSON20131234}.
P.B.L. was supported by the National Natural Science
Foundation of China under Grant No. 12375018. X.L.D. was supported by the National Scholarship Foundation of China under Grant No. 202206280115.
Z.G. was supported by The University of Tokyo Excellent Young Researcher Program.
F.N. is supported in part by:
the Japan Science and Technology Agency (JST)
[via the CREST Quantum Frontiers program Grant No. JPMJCR24I2,
the Quantum Leap Flagship Program (Q-LEAP), and the Moonshot R\&D Grant Number JPMJMS2061],
and the Office of Naval Research (ONR) Global (via Grant No. N62909-23-1-2074).


%

\end{document}